\documentclass[preprint2]{aastex}
\usepackage{amsmath}
\begin{document}

\title{Detectability of GRB Optical Afterglows with $\textit{Gaia}$ Satellite}

\author{J. Japelj$^1$ and A. Gomboc$^{1,2}$}

\affil{$^1$Faculty of Mathematics and Physics, University of Ljubljana, Jadranska ulica 19, SI-1000 Ljubljana, Slovenia \\
	      $^2$ Centre of Excellence SPACE-SI, A\v sker\v ceva cesta 12, SI-1000 Ljubljana, Slovenia}

\affil{e-mail:  jure.japelj@fmf.uni-lj.si, andreja.gomboc@fmf.uni-lj.si}

\begin{abstract}
With the launch of the $\textit{Gaia}$ satellite, detection of many different types of transient sources will be possible, with one of them being optical afterglows of gamma-ray bursts (GRBs). Using the knowledge of the satellite's dynamics
and properties of GRB optical afterglows, we performed a simulation in order to estimate an average GRB detection rate with $\textit{Gaia}$. Here, we present the simulation results for two types of GRB optical afterglows, differing in the observer's line of sight compared with a GRB jet axis: regular (on-axis) and orphan afterglows. Results show that for on-axis GRBs, less than 10 detections in five years of foreseen $\textit{Gaia}$ operational time are expected. The orphan afterglow simulation results are more promising, giving a more optimistic number of several tens of detections in five years.  
 \end{abstract}


\section{Introduction}
\label{s1}
European Space Agency's cornerstone mission $\textit{Gaia}$ is scheduled for launch in 2013. It is primarily an astrometric mission, designed to measure precise 3D positions for up to a billion stars in our Galaxy and other bright objects in the sky. In addition to astrometric measurements, spectroscopy and multi-band photometry will be performed, providing radial velocities and additional astrophysical information on the stars and other objects. The main scientific goal of $\textit{Gaia}$ mission is to clarify the origin and formation history of our Galaxy.  The goals of the mission are described in \citet{perryman}, \citet{lindegren}, and, more thoroughly, in \citet{turon}.

$\textit{Gaia}$ will continuously scan the sky, covering the whole sky and building its own catalogue of celestial objects. A preliminary catalogue will be released every 6 months and a final catalogue, exploiting all the data recovered during the mission, will be released at the end of the operational time (around 2020) \citep{lammers}. Comparing further observations with this catalogue will enable $\textit{Gaia}$ to detect various transient sources, with some of them being cataclysmic variables, supernovae, active galactic nuclei, asteroids, etc. Since the catalogue of $\textit{Gaia}$ astrometric and photometric data will be available only late into the $\textit{Gaia}$ operation, triggers from transient phenomena will actually be the first $\textit{Gaia}$ data released to the scientific community. In order to be prepared for this stream of data and to make sure the alert stream is accurate, reliable, and free from contamination, the $\textit{Gaia}$ Science Alerts Working Group has been formed\footnote{http://www.ast.cam.ac.uk/ioa/research/gsawg}. 

One type of possible transients to be detected by $\textit{Gaia}$ is gamma-ray burst optical afterglows. Gamma-ray bursts (GRBs) are extremely energetic explosions occurring at cosmological distances \citep{piran,meszaros}. They represent the most luminous events in the gamma-ray part of the electromagnetic spectrum known in the universe. Usually, the short-term prompt gamma-ray emission is followed by an afterglow emitted at longer wavelengths from X-rays to radio waves, which can last for several days. According to the standard fireball model \citep{reesa, reesb} the prompt gamma emission is produced in the internal shocks of the relativistically expanding ejecta, while the afterglow is synchrotron emission from relativistically accelerated electrons in the external shocks, created in the ejecta interaction with the circumburst medium. In general, afterglow light curves exhibit the power-law decay with time \citep{sari}, although many additional features like density bumps \citep{lazzati,guidorzi}, late energy injections \citep{nousek,evans} and reverse shocks \citep{akerlofc,fox,kobayashi,gombocb} have been seen in recent years, when better time coverage observations became possible due to accurate GRB alerts from the $\textit{Swift}$ satellite \citep{gehrels}  and a number of follow-up robotic telescopes (e.g., the Liverpool Telescope \citep{gomboc}, ROTSE \citep{akerlofa}, REM \citep{zerbi}, etc.).

GRB explosions are believed to be collimated, rather than spherical \citep{rhoadsa,granot}. An indirect argument supporting this is based on the high values of energy output in gamma rays ($E_{\rm iso}$), which is of the order of $\sim M_{\odot}c^{2}$ if we assume that the explosion is spherically symmetric. It is known that GRBs occur in galaxies, and are of stellar origin (most probably the collapse of massive, rapidly rotating stars and mergers of compact objects). It is difficult to conceive  a model with a stellar progenitor, which can produce such high energies inferred in the case of spherically symmetric explosions. More direct evidence comes from precise observations of afterglow light curves: an achromatic  break in the light curve, which has been observed in many GRB afterglows at a few hours to days after the GRB itself (\citealt{granot} and references therein), was predicted theoretically prior to observational discovery \citep{rhoads,sarib}. The break (usually referred to as a jet break) can be well explained by considering a jet with a half-opening angle $\theta_{\rm j}$, moving initially at a relativistic speed with high Lorentz factor $\Gamma \sim 100 - 1000$. Because of the relativistic beaming effect, the radiation emitted from the jet is beamed in a half-opening angle $\theta_{\rm b} = \frac{1}{\Gamma}$ in the direction of motion. The expanding jet is surrounded with a medium that causes the jet to decelerate. This results in an increasing value of the angle $\theta_{\rm b}$. Shortly after the prompt GRB emission, we cannot distinguish between a spherical and collimated explosion due to strong beaming, since $\theta_{\rm j} > \theta_{\rm b}$. But after some time, the jet slows down and $\theta_{\rm b}$ becomes larger than the opening angle of the jet itself. At this point, the radiation flux drops and we can see an achromatic break in the light curve \citep{kulkarni,klose}. As mentioned previously, a collimated explosion results in a reduced estimation of energy output, i.e., $E_{\rm j} = \frac{1}{2}\theta_{\rm j}^{2}E_{\rm iso}$, where $E_{\rm j}$ is collimation-corrected energy in the case of a jet with half-opening angle $\theta_{\rm j}$.

The shape of the afterglow light curve depends on the relative position of the observer's line of sight and the axis of the jet cone. We can describe that with an angle $\theta_{\rm obs}$ between both directions. It is very unlikely for an observer to be looking precisely at the direction of the jet axis (to have $\theta_{\rm obs}=0$). We get an 'on-axis' GRB and a 'regular' optical afterglow when $\theta_{\rm obs} < \theta_{\rm j}$. If the jet cone is directed so that $\theta_{\rm obs} > \theta_{\rm j}$,  an observer, due to the beaming effect, probably cannot observe the prompt gamma emission, i.e., detect a GRB. But the deceleration of the jet will gradually expand the beaming angle and the observer might detect the afterglow. The observed afterglow that was not preceded by a prompt gamma emission is called an orphan afterglow (OA)\footnote{The OAs we are discussing here are not to be confused with optical afterglows, for which the observer does not detect the prompt gamma emission for some reason, even though the jet cone is directed into the observer's line of sight.}. Even though OAs have not been conclusively detected yet, they have been studied a great deal over the last 15 years (\citealt{granot} and references therein). 

As part of the scientific community preparation for the $\textit{Gaia}$ mission, we discuss $\textit{Gaia}$'s potential for detecting these short, rapidly fading and unpredictable transients. We present here results of the simulation we performed in order to estimate the number of possible GRB optical afterglow detections with the $\textit{Gaia}$ satellite. The simulation is divided in two parts, the first one concerning on-axis GRB afterglows and the second one dealing with orphan afterglows. In Section 2 we briefly discuss the dynamics and basic characteristics of $\textit{Gaia}$. The simulation with all the necessary parameters for both on-axis and orphan afterglows is described in Section 3. In Section 4, we present and discuss results of the simulation and possible ways of afterglow identification. 

\begin{figure}[!t]
\epsscale{1}
\plotone{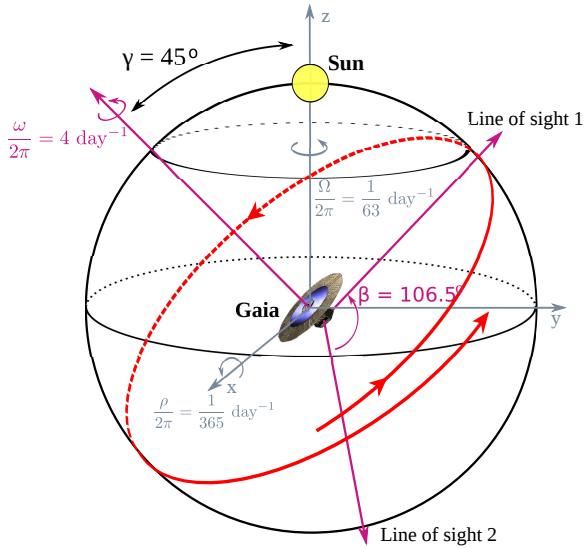}
\caption{Sketch showing how $\textit{Gaia}$ will scan the sky. The telescope axes are separated by an angle of $\beta = 106.5^{\circ}$. The satellite spins with a constant angular velocity $\omega$ around its main axis, which is kept at an angle $\gamma = 45^{\circ}$ from the Sun at all times. The main axis experiences slow precessional motion around the Earth-to-Sun direction ($\Omega$). Rotation around the Sun (angular velocity $\rho$) is also taken into account. Details of the dynamics are given in the Appendix.}
\label{fig1}
\end{figure}

\section{The $\textit{Gaia}$ Satellite}
\label{s2}

For the purpose of our simulation, we need to know the way $\textit{Gaia}$ will scan the sky \citep{lindegren}. $\textit{Gaia}$ will carry two identical telescopes, separated by the angle of $\beta = 106.5^{\circ}$, as shown in Figure \ref{fig1}. The satellite will make  four rotations per day around its axis (which is perpendicular to the direction  in which both telescopes are pointing) with the constant angular velocity $\omega$. The direction of the axis itself is tilted by the angle $\gamma = 45^{\circ}$ from the direction of the Sun. The axis will experience slow precession motion around the Earth-to-Sun direction with a period of 63 days ($\Omega$ in Figure \ref{fig1}). $\textit{Gaia}$ will have an orbit around the L2 point and will thus experience rotation around the Sun, which is shown in Figure \ref{fig1} as a rotation around the $x$ axis. Knowing $\omega, \Omega, \gamma, \beta$ and $\rho$ (one year orbital period around the Sun), we construct the scanning law of $\textit{Gaia}$ (see the Appendix \ref{a1}). 

$\textit{Gaia}$'s two telescopes will have a field of view of $\sim 0.7^{\circ}\times 0.7^{\circ}$ each. The expected limiting magnitude in broadband $G$ magnitude (details on the photometric system of $\textit{Gaia}$ are given in \citealt{jordi}) is $G$ = 20 mag. The specifics of scientific performance are given on the official $\textit{Gaia}$ World Wide Web page\footnote{http://www.rssd.esa.int/index.php?project=$\textit{Gaia}$\& page=index} and, for example, in \citet{lindegren}. There will be a 7 $\times$ 9 astrometric CCD field in $\textit{Gaia}$'s focal plane. Each source will transit over the nine CCDs and will be observed by each of them  with 4.4 s integration time. 

\section{The simulation}
\label{s3}

\subsection{On-Axis Afterglows}
\label{ss1}
First we focus on on-axis afterglows, i.e., those with their jet cones turned in our line of sight. In such cases we can observe the GRB, which triggers a satellite and follow-up optical observations.  Hence, we can base the initial parameters of our simulations on the actually observed GRB afterglow numbers and their characteristics. Since the launch of the $\textit{Swift}$ satellite in 2004 \citep{gehrels} a large number of GRBs and their afterglows has been detected \citep{roaming}. The $\textit{Swift}$ detection rate is about 100 GRBs per year. In about half of the detected GRBs there is no bright optical afterglow detected. Since the satellite covers approximately $\frac{1}{6}$ of the sky, we can estimate that there are around 300 GRBs per year, for which an optical afterglow could be detected with timely observations. To obtain their general properties, we used observations published in the Gamma Ray Burst Coordinate Network Circulars\footnote{http://gcn.gsfc.nasa.gov/} \citep{barthelmy}. We chose GRBs detected between 2006 September and 2009 April that had an optical afterglow detected (and reasonably well sampled). In general, not many optical afterglows have been detected in the first few minutes after the initial trigger \citep{kanna,kannb}. Consequently, we used measured $R$ magnitudes at approximately $(t-t_0)=$0.01 day after the prompt GRB ($t_0$ is the time of the GRB trigger), at which the number of detected afterglows is larger; if an afterglow was first detected at a later time, we estimated its $R$ magnitudes at $(t-t_0)=$0.01 day from its light curve. Our sample includes 100 GRBs, with their optical afterglow magnitudes at $(t-t_0)=$0.01 day spanning a range of values between $R=13$ mag and $R=23$ mag.

\begin{figure}[!t]
\epsscale{1}
\plotone{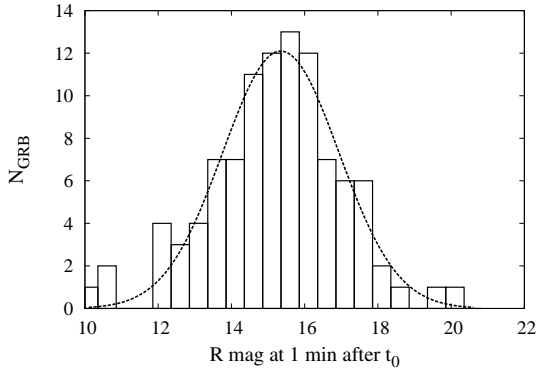}
\caption{Number of observed GRB afterglows vs. their $R$ magnitudes shifted to 1 minute after the GRB trigger. The distribution is fitted with a normal function with mean value $\mu = 15.35 \pm 0.09$ and standard deviation $\sigma = 1.59 \pm 0.09$. Numbers are based on data from GCN circulars.}
\label{fig2}
\end{figure}

In general, GRB optical afterglows follow a power-law decay, $F_{\nu} (t) \propto t^{-\alpha}$, with typical values of temporal decay index $\alpha$ scattered between 0.4 and 1.4 \citep{akerlofb}. These values correspond to the time before a jet break, which is evident in some, but not all afterglows. To model the afterglows from early time on (where observations are still rare), we shift the distribution of $R$ magnitudes at $(t-t_0)$=0.01 day to $(t-t_0)$=1 minute, using the average temporal index $\alpha = 1.0$. The time scale of 1 minute usually corresponds to the onset of afterglow emission\footnote{There are examples of prompt optical emission in the first minute after a trigger but they are scarce.}. We somehow arbitrarily chose the time 1 minute after the GRB as the start of an afterglow in order not to miss any possible detection. This choice does not alter our results, since, as will be shown in Section \ref{ss3}, the probability of afterglow detection in the first 0.01 day after the burst is small and, therefore we could have chosen any time between 1 minute and 0.01 day. We find that the distribution of magnitudes at $(t-t_0)$=1 minute is well described by the normal distribution with the mean value $\mu = 15.35 \pm 0.09$ mag and standard deviation $\sigma = 1.59 \pm 0.09$ mag. The distribution of $R$ magnitudes, shifted to $(t-t_0)$=1 minute, and corresponding normal function are shown in Figure \ref{fig2}.

The temporal and sky distribution of GRBs is random, with about 300 GRBs per year, for which an optical afterglow could be detected. Following that, in our simulation we generate $300\times 5$ GRB events distributed randomly (but uniformly) over the sky and randomly (but uniformly) in time (i.e., random $t_0$) over the five year period. Magnitudes of generated GRB afterglows at  $(t-t_0)$=1 minute are distributed according to the aforementioned normal distribution. Next we assume that the magnitudes are, assuming $\alpha = 1.0$, evolving with time as:
\begin{equation}
\label{eq2}
m(t-t_0) = m(1 ~\rm min) + 2.5\log \left( \frac{t-t_0}{1 ~\rm min}\right).
\end{equation}
Since the distribution of GRBs is random both in space and time, we do not have to specify a special initial position of the telescope's axes. We simply put the first telescope (number 1 in Figure \ref{fig1}) at the origin of a fixed coordinate system at time $t = 0$ and from there on compute the position of the telescope's axis at time $t  >  0$ according to the scanning law.
After we generate a GRB, we first calculate the time $t_{\rm lim}$ after which the GRBs magnitude will fall below the limiting magnitude. We then check where the positions of both telescopes are at $(t-t_0)$=1 minute, i.e., if they are pointing in the direction of a GRB. If they are not, we calculate the telescopes' axes paths over the sky. We stop the simulation when the time $t_{\rm lim}$ expires and check if there were any detections with the first or second telescope. After that, the axes of telescopes are returned to the initial position, another GRB is generated, and the described procedure is repeated.

\subsection{Orphan Afterglows}
\label{ss2}
The simulation in this part is done in a similar way as in the case of on-axis afterglows. Again, we generate a number of events, distributed randomly (but uniformly) in the sky and randomly (but uniformly) in time. The difference from the preceding section is that  the OAs have not been conclusively observed yet, and therefore we do not have any observational data available on which to base the number of observable OAs and their light curve properties. We approach the problem by constructing a number of OAs' light curves with characteristics from theoretical predictions. 

Several semianalytical models have been proposed to explain the dynamics of the jet and, consequently, the shape of the OAs' light curve \citep{granot}. In our simulation, we follow the approach of \citet{zou}, which in itself is based on \citet{wu} and \citet{granotb}, where an adiabatic jet with a half-opening angle $\theta_{\rm j}$ and without sideways expansion is considered. The latter is a good approximation for a highly relativistic jet, as have been shown in hydrodynamical simulations \citep{granot}.

With the knowledge of the jet half-opening angle $\theta_{\rm j}$, its total collimation-corrected energy $E_{\rm j}$, redshift $z$, and the density of the circumburst medium $n$, we can compute the time $t_{\rm j}$ at which an on-axis observer will see the break in a light curve and the flux density $F_{\nu,\rm j} = f(t_{\rm j}, E_{\rm j}, D_{\rm L}, z, \nu, p, \varepsilon_{\rm e}, \varepsilon_{\rm B})$ at that time (equations [2] and [3] in \citealt{zou}); $D_{\rm L}$ stands for luminosity distance, $p$ is the power-law index of shock-accelerated electrons, and $\varepsilon_{\rm e}$ and $\varepsilon_{\rm B}$ are the energy equipartition factors of the electrons and magnetic field, respectively. The rest of the light curve is obtained by applying the power-law temporal decay with a prebreak temporal index $\alpha_{1} = 1.0$ (the same as in the on-axis case) and a postbreak index $\alpha_{2} = \alpha_{1} + 3/4$ \citep{zou}. 

\begin{figure}[!t]
\epsscale{1}
\plotone{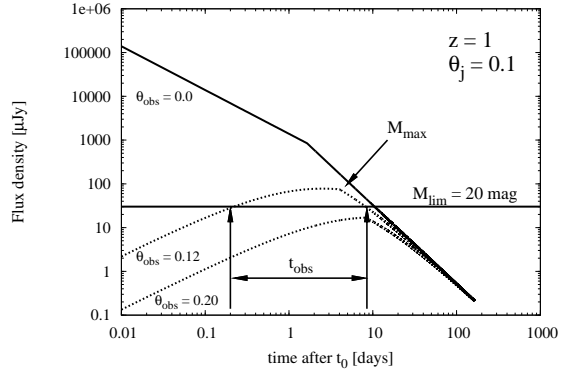}
\caption{Example of an optical afterglow light curve ($\nu = 4.55 \times 10^{14}$ Hz) seen by an on-axis observer (solid line) and the same event seen by an off-axis observer at an angle $\theta_{\rm obs} = 0.12$ and $0.20$ (dotted lines). $t_{0}$ is the time at which the burst occurred.}
\label{fig3}
\end{figure}

The light curve seen by an off-axis observer is obtained by considering a point-source approximation \citep{granotb}:
\begin{equation}
\label{eq3}
F_{\nu}(\theta_{\rm obs}, t) = a^{3}F_{\rm \nu/a}(0, at),
\end{equation}
where $a \equiv (1 - \beta)/(1 - \beta \cos \theta_{\rm obs})$ and $\beta = \sqrt{1 - 1/\Gamma^{2}}$. This approach works only in the case of $\theta_{\rm obs} > \theta_{\rm j}$. Time evolution of the Lorentz factor is described by \citep{granotb}:
\begin{equation}
\label{eq4}
\Gamma(t) = \left \lbrace
\begin{array}{ll}
\theta_{\rm j}^{-1}\left( \frac{t}{t_{\rm j}}\right)^{-3/8} & t < t_{\rm j}\\
\theta_{\rm j}^{-1}\left( \frac{t}{t_{\rm j}}\right)^{-1/2} & t > t_{\rm j}\\
\end{array}
\right.
\end{equation}
In Figure \ref{fig3} three afterglow light curves are shown for the case of $\theta_{\rm j} = 0.1$ and $z = 1$. The flux is calculated for an observed frequency $\nu = 4.55 \times 10^{14}$ Hz. The light curve drawn with full line corresponds to an on-axis observer, while the dotted lines represent the light curves seen at angles $\theta_{\rm obs} = 0.12$ and $0.20$. The lightcurves of OAs cannot be observed immediately after the GRB. However, the flux slowly rises and eventually reaches the peak value, after which it begins to decay again. Depending on the parameters mentioned in this section, the OA's light curve can be seen during the period in which the OA's flux is above the observer's detection limit.

In our simulation, a light curve for a given set of parameters is calculated. It is then verified, whether it rises above the limiting flux or not. In the former case, the time interval $t_{\rm obs}$ in which the OA can be observed is calculated. If during this time interval the position of the burst comes into the field of view of one of the $\textit{Gaia}$ telescopes, we consider that the afterglow is detected.

Parameter values used in our simulation are: $E_{\rm j} = 1\times 10^{51}$ ergs, $p = 2.2$, $\varepsilon_{\rm e} = 0.1$, $\varepsilon_{\rm B} = 0.01$, and $\nu = 4.55 \times 10^{14}$ Hz. We assign the same value of $E_{\rm j}$ to all generated bursts, since it is believed that different observed energy values correspond to different jet half-opening angle values \citep{zou}, and we do not expect that this will have substantial effect on our results. Other parameters need a more detailed discussion.

\subsubsection{Number of GRBs}
\label{sss1}
Several estimations of the detectability of OAs have been made \citep{zou,totani,nakar}. Their results depend on jet models, initial half-opening angle distribution, medium surrounding the GRB source, etc. However, we decided to make a simple estimation of the number of all GRBs that occur in the universe up to some redshift $z$, as described subsequently. During the simulation, the events with $\theta_{\rm obs} < \theta_{\rm j}$ are sorted out and we are left only with the bursts for which OA light curves can be generated.

Following \citet{guetta} the GRB formation rate is approximated to follow the Rowan-Robinson star formation rate \citep{rowan} 
\begin{equation}
\label{eq5}
R_{\rm GRB} = \rho_{0} \left\lbrace
\begin{array}{ll}
10^{0.75z} & z < 1\\
10^{0.75} & z \geq 1,
\end{array}
\right.
\end{equation}
where $\rho_{0} \sim 33 ~h_{65}^{3}\rm ~Gpc^{-3}~yr^{-1}$ \citep{guetta}. Considering the expansion of the universe (which is taken to be flat, i.e., $\Omega_{k} = 0$), the number of bursts $dN$ that occur between redshift $z$ and $z + dz$ in one year is:
\begin{equation}
\label{eq6}
dN(z) = \frac{R_{\rm GRB}}{1 + z}\frac{dV}{dz}dz,
\end{equation}
where the division factor $(1 + z)$ is due to time dilation. The comoving volume element $dV/dz$ equals
\begin{equation}
\label{eq7}
\small
\frac{dV}{dz} = \frac{c}{H_{0}}\frac{4\pi \chi^{2}(z)}{E(\Omega_{\rm i}, z)}, \hspace{0.3cm} \chi(z) = \frac{c}{H_{0}}\int_{0}^{z} \frac{dz^{'}}{E(\Omega_{\rm i},z^{'})}
\end{equation}
with $E(\Omega_{\rm i}, z) = \sqrt{\Omega_{\rm m}(1 + z)^{3} + \Omega_{\Lambda}}$. The number of GRBs per year is:
\begin{equation}
\label{eq8}
N = \int_{0}^{z_{\rm max}} \frac{R_{\rm GRB}}{1 + z}\frac{dV}{dz}dz.
\end{equation}
In our calculation, we adopt standard cosmology of a flat universe with $h = 0.72$, $\Omega_{m} = 0.3$, and $\Omega_{\Lambda} = 0.7$ and set $z_{\rm max} = 5$. Thus, the calculated number of GRBs that occur per year is $N \sim 25000$.

\subsubsection{GRB Redshift Distribution}
\label{sss2}
We generate the redshift distribution of GRB events according to equation (\ref{eq6}) in the following way. The probability of finding a GRB at a redshift $z$ is \citep{coward}:
\begin{equation}
\label{eq9}
p(z) = \frac{(dN/dz)}{\int_{0}^{z_{\rm max}} (dN/dz)dz }.
\end{equation}
Next, a cumulative function $P(z)$ can be calculated as: 
\begin{equation}
\label{eq10}
P(z) = \int_{0}^{z} p(z^{'})dz^{'},
\end{equation}
giving the probability of an event occurring in the redshift range from $z=0$ to $z$. By calculating the inverse of $P(z)$, the redshift as a function of the cumulative probability is obtained. Employing a random number generator to select values $P(z)$, a GRB redshift distribution according to equation (\ref{eq6}) is generated. 

\subsubsection{Jet Half-Opening Angle Distribution} 
\label{sss3}
Not many GRB afterglows have been observed with a high enough precision to enable us the determination of $\theta_{\rm j}$, which can be calculated as \citep{frail}
\begin{equation}
\small
\begin{split}
\label{eq11}
\theta_{\rm j} &= 0.057\left( \frac{t_{\rm j}}{1 ~\rm day}\right)^{3/8}\left( \frac{1 + z}{2}\right)^{-3/8}\cdot \\
&\cdot \left( \frac{E_{\rm iso}(\gamma)}{10^{53}~\rm erg}\right)^{-1/8}\left( \frac{n_{\gamma}}{0.2}\right)^{1/8}\left( \frac{n}{0.1 ~\rm cm^{-3}} \right)^{1/8},
\end{split}
\end{equation}
where $E_{\rm iso}(\gamma)$ is the gamma energy output assuming a spherical burst, $n_{\gamma}$ is the efficiency in converting the energy of the ejecta into gamma rays and $n$ is the number density of the circumburst medium. 

\begin{figure}[!t]
\epsscale{1}
\plotone{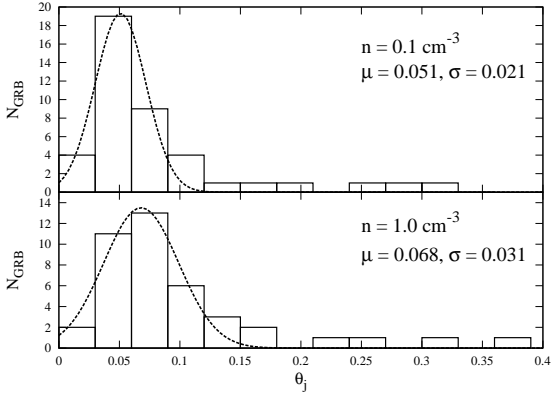}
\caption{Distribution of calculated jet half-opening angles $\theta_{\rm j}$ for two different values of $n$. In both cases the distribution is fitted with a normal distribution with the mean value $\mu$ and standard deviation $\sigma$.}
\label{fig4}
\end{figure}

Using equation (\ref{eq11}), we calculated angles $\theta_{\rm j}$ for 43 GRBs from observational data collected in Table 1 in \citet{gao}. The obtained angle distribution is shown in Figure \ref{fig4}. In our calculation, we set $n_{\gamma} = 0.2$. We chose two different values of number density and ran the simulation for each number density case separately. The angle distributions have been fitted with a normal distribution function, and in the simulation, randomly generated half-opening angles are taken to be distributed according to these two functions. As an additional constraint we generate only angles greater than $\theta_{\rm j} = 0.01 ~\rm rad$.   

Another parameter in the simulation is the observational angle $\theta_{\rm obs}$, which is distributed uniformly between 0 and $\pi$ (due to jet bimodality).

\section{Results and discussion}
\label{s4}
Here, we present results for both parts of the simulation. All results are given for a time period of five years, which is the expected operational time of $\textit{Gaia}$. All results are averaged over 1000  and 100 simulations for on-axis and OA, respectively. 

\begin{figure}[!t]
\epsscale{1}
\plotone{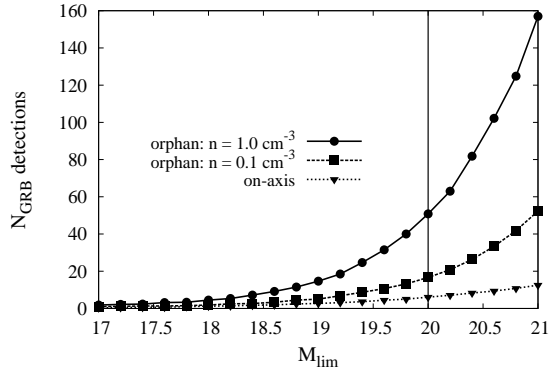}
\caption{GRB on-axis and OA detection rate in five years of $\textit{Gaia}$ operation as a function of $M_{\rm lim}$. In the case of OAs, the results are given for two different circumburst environments.}
\label{fig5}
\end{figure}

Figure \ref{fig5} shows the number of GRB optical afterglows detected by $\textit{Gaia}$ in five years depending on the value of $\textit{Gaia}$'s limiting magnitude. Results for the case of the on-axis simulation are not very promising. In the nominal $\textit{Gaia}$ $M_{\rm lim} = 20$ mag, we can expect to see less than 10 on-axis afterglows in five years. Possible detection of OAs is more promising: with $M_{\rm lim} = 20$ mag, we expect to detect several tens of OAs in five years. The latter is actually an upper limit:  possible light extinction in the line of sight in the Galaxy and GRB host galaxies \citep{schady} has not been included in the calculation of the OAs' light curves. The results of the on-axis simulation, which is based on observational data, already include the extinction effect. Also (and this applies to both on-axis and OAs), the simulation was done for the case of the regular Johnson-Cousins $R$ magnitude \citep{bessel}, while the $\textit{Gaia}$ $G$ magnitude is obtained with a filter of a much wider passband \citep{jordi}. This results in an overall slightly fainter limiting magnitude.

Hereafter, all results are given for $M_{\rm lim} = 20$.

\begin{figure*}[!t]
\epsscale{2.3}
\plottwo{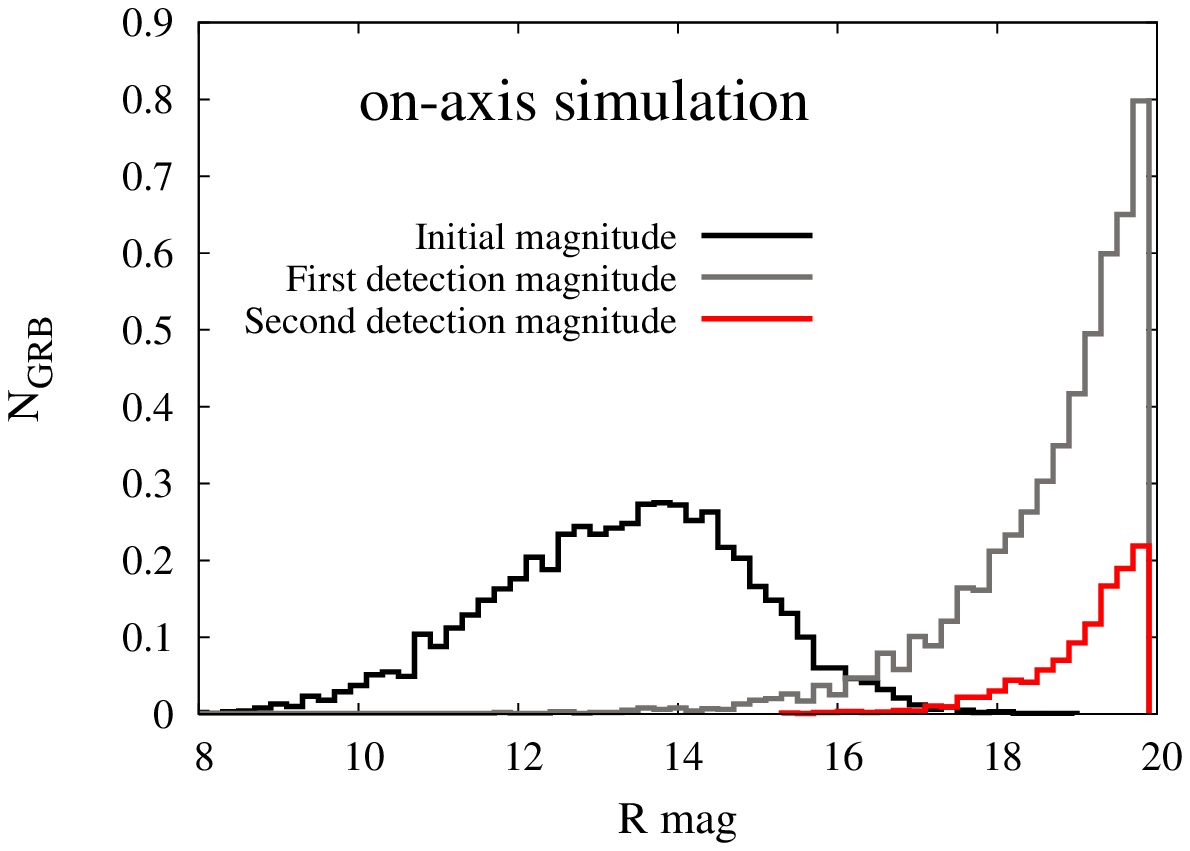}{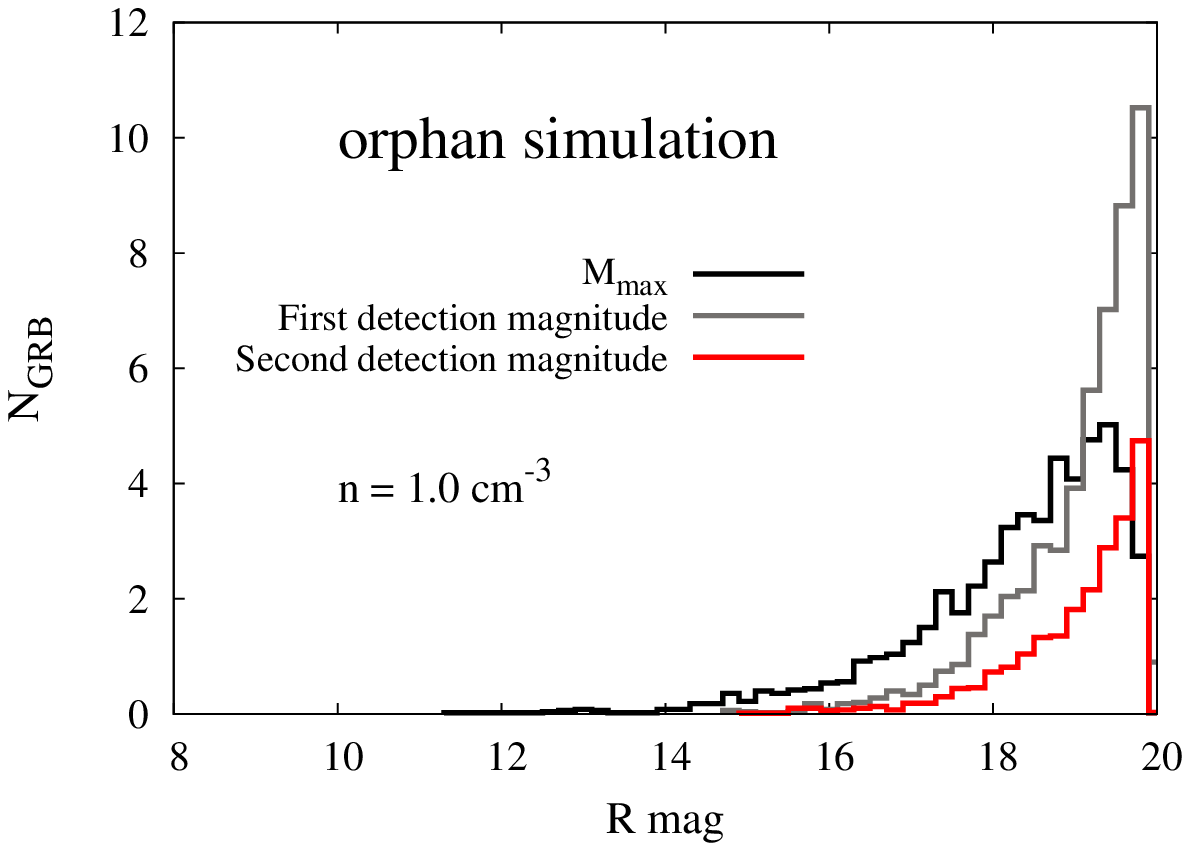}
\caption{Distribution of magnitudes that are associated with detected afterglows (black), same magnitudes at the time of the first detection (gray), and magnitudes of afterglows detected by the second telescope (red) in the case of on-axis (left) and OA (right) simulation. The results are given for a time of five years.}
\label{fig6}
\end{figure*}

For the on-axis case, the distribution of initial magnitudes of optical afterglows, which were detected by (at least) one of the $\textit{Gaia}$ telescopes, is given in  Figure \ref{fig6} (left). It is a normal distribution with the mean value $\sim 13.5$ mag. This was expected, since the afterglows with brighter initial magnitudes are more likely to be detected (their $t_{\rm lim}$ is longer). It is interesting to see the distribution of magnitudes at the time of their first detection (gray distribution in Figure \ref{fig6}). Since the majority of detected afterglows are already quite faint, we cannot expect many of them to be detected by the second telescope. Indeed, the distribution of twice-detected afterglows shown in Figure \ref{fig6} in red tells us that, given the small number of detections with one telescope, the probability to detect an afterglow with both telescopes is low. A similar plot was made for the OA simulation and is shown in Figure \ref{fig6} (right). The distribution of magnitudes in the case of $n = 0.1 \rm ~cm^{-3}$ is not given here, since it strongly resembles that of $n = 1.0 \rm ~cm^{-3}$ (but with the lower number of detections). The fraction of twice-detected OAs is larger here. The decay in the on-axis case is constant with $\alpha = 1$, while in the OA case the flux could actually still be rising at the time of first detection. In addition, the flux decay around $M_{\rm max}$ is much shallower than $\alpha = 1$, which explains the larger fraction of second detections in OA case.

\begin{figure*}[t!]
\epsscale{2.3}
\plottwo{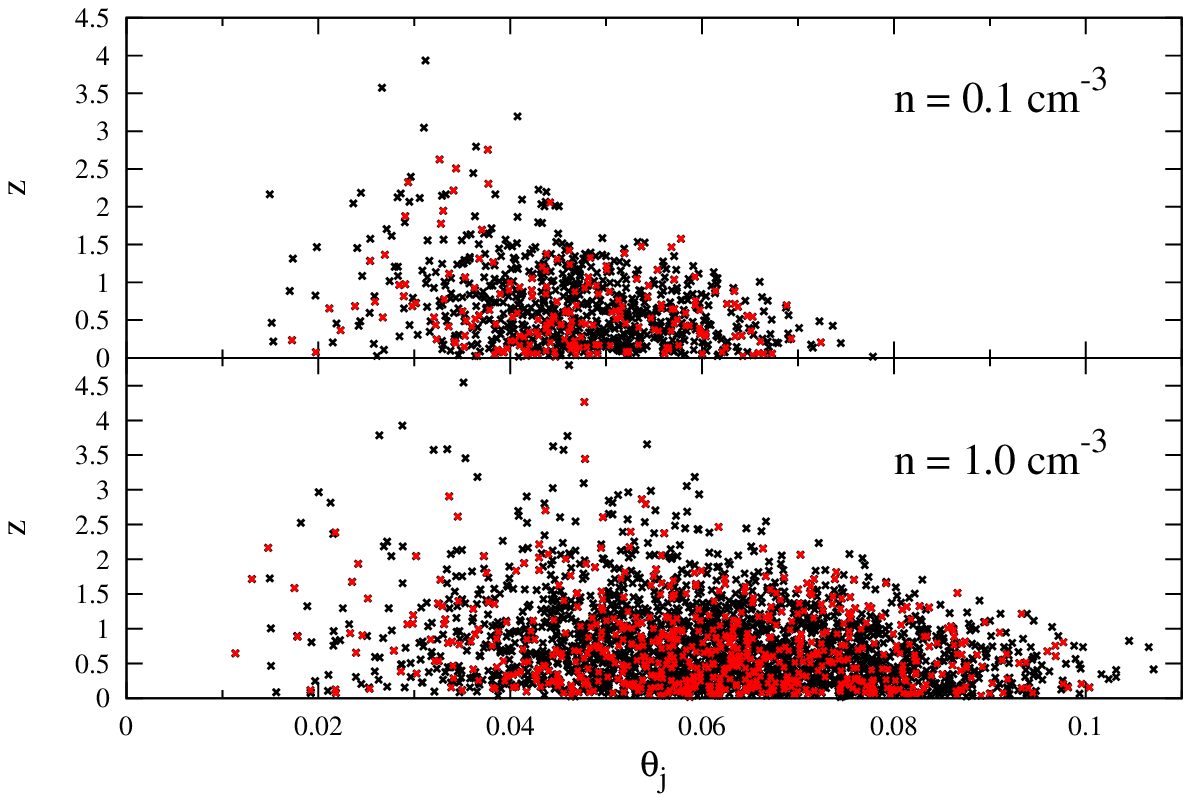}{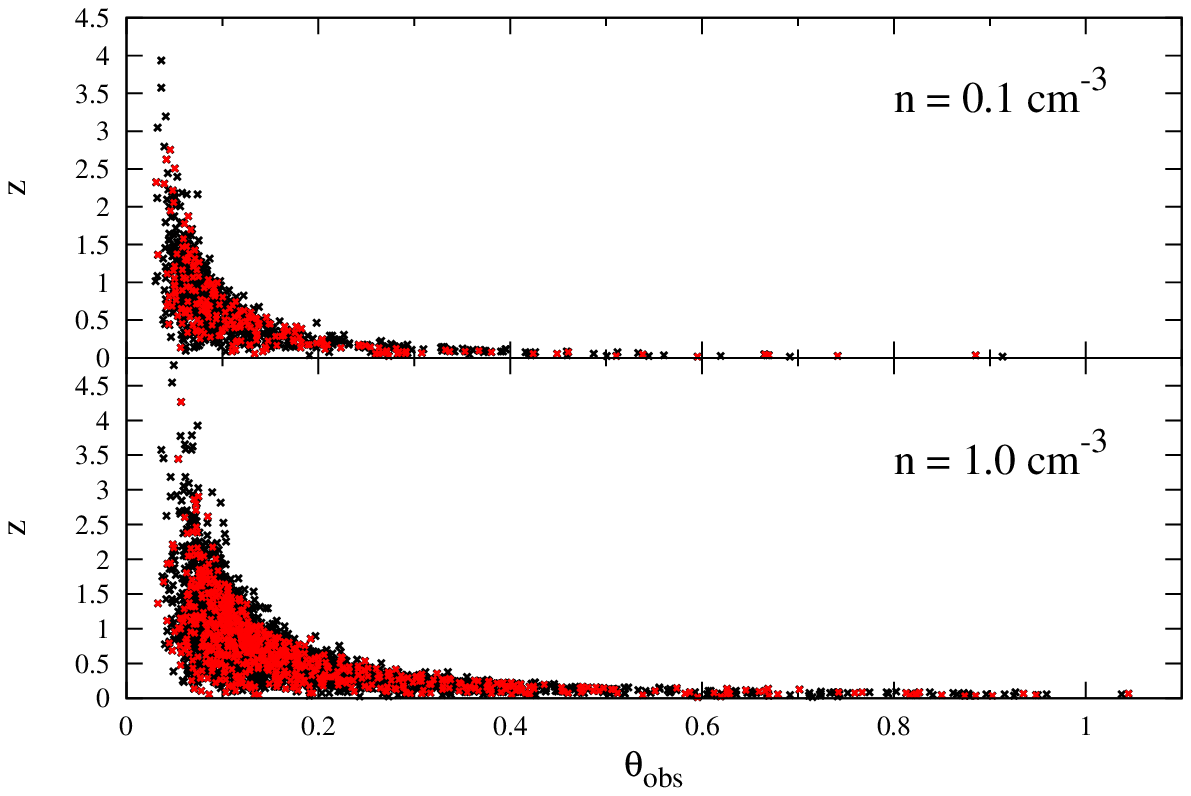}
\caption{Correlation between redshift $z$ and jet half-opening angle $\theta_{\rm j}$  (left) and observing angle $\theta_{\rm obs}$ (right) for detected OAs. Black points represent OAs detected with one telescope, and red points represent OAs detected by both telescopes.}
\label{fig7}
\end{figure*}

To check the sensibility of OA simulation results, we look at the correlation between redshift $z$ and jet half-opening angle $\theta_{\rm j}$, shown in Figure \ref{fig7} (left). As expected, most $\theta_{\rm j}$ values are concentrated around the peak of $\theta_{\rm j}$ distribution (section \ref{sss3}). Most points have values of $z < 2$. This can be understood, since higher redshift value results in a lower observed flux. In addition, the GRB redshift distribution (section \ref{sss2}) peaks at $z = 1$ and starts to decline at higher $z$. Nevertheless, a few OAs with high redshift are detected in the simulation. Since all of these OAs have relatively low $\theta_{\rm j}$, which neutralizes the negative redshift effect on the flux value, their detection is understandable. The majority of second detections have low redshift values and no preferable $\theta_{\rm j}$. This is mostly because the majority of detections with the first telescope have low redshift values. Because of the relatively flat light curve around the peak value (Figure \ref{fig3}), once the OA has been detected with the first telescope, redshift and $\theta_{\rm j}$ are not the crucial parameters for the second detection. In order to detect OA with the second telescope, the time $t-t_0$ of the first detection, i.e., before or after the peak flux value, is important.

Figure \ref{fig7} (right) also shows the correlation between redshift $z$ and observing angle $\theta_{\rm obs}$ (which for the OAs has to be larger than $\theta_{\rm j}$ of a particular burst). If a burst has a low redshift, $\theta_{\rm obs}$ can be quite large, as is evident in the long tail. On the other hand, a very low observing angle helps in the detection of an OA from a distant object, since the difference between $\theta_{\rm j}$ and $\theta_{\rm obs}$ is small and the burst is 'almost' on-axis. Detections with the second telescope do not appear to have a preferable position in the graph.

\subsection{Identification of GRB Optical Afterglows}
\label{ss3}

\begin{figure}[t!]
\epsscale{1}
\plotone{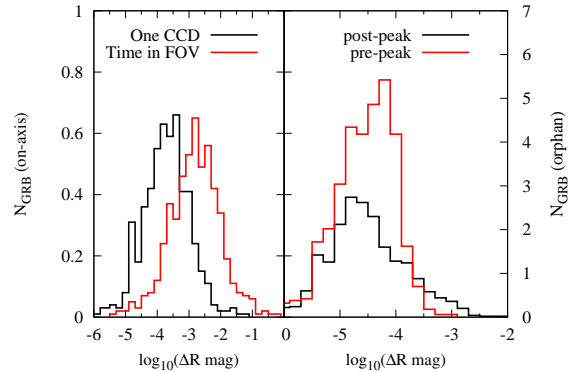}
\caption{Left: Distribution of the change in magnitude in 4.4 s (corresponding to an observation with two successive CCDs) (black) and the change of magnitude in the time in which an afterglow is in the entire field of view (red) for the on-axis case. Right: Distribution of the change in magnitude in the time in which an orphan afterglow is in the field of view. Black distribution corresponds to detection after the peak in a light curve, and red distribution corresponds to detection before the peak.}
\label{fig8}
\end{figure}

On-axis afterglows are initially rapidly fading and they will be observed for 4.4 s by each of nine transiting CCD detectors. In principle, the change of magnitude in that time, i.e., in observations by two successive CCDs or CCD field transit, could be observable. Therefore, we calculated the change of magnitude in 4.4 s for all detected on-axis afterglows and also for observations lasting as long as a particular afterglow is in the telescope's field of view. The distribution of the change of magnitude for both cases is shown in Figure \ref{fig8} (left). We calculated the changes only for the first afterglow detection, since the probability of detection with the second telescope is small. In practice, the change of magnitude could be detected only if it was larger than the photometric error. The expected photometric errors, averaged across the sky, in the case of $M_{\rm lim} = 20$ mag will be 10 mmag at best (though it will be smaller for brighter events) \citep{perryman}. The chances of observing the change in magnitudes will thus be limited. The change in magnitude of OAs will be even harder to detect, since the flux rises and decays more slowly. Also, OA could actually become brighter through the observation if the detection happened prior to the peak light-curve value (Figure \ref{fig8}, right).

\begin{figure}[t!]
\epsscale{1}
\plotone{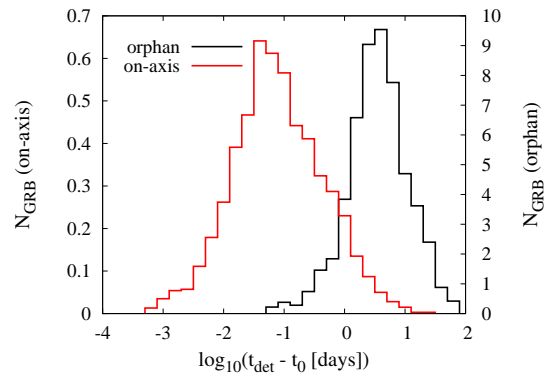}
\caption{Distribution of first detection times $t_{\rm det}$ relative to the time $t_{0}$  of the initial GRB for the case of on-axis (red) and orphan afterglows (black).}
\label{fig9}
\end{figure}

Looking at the time of detection relative to the initial GRB (Figure \ref{fig9}), most of on-axis afterglows are expected to be detected at $\sim 0.1$ day after the GRB. The time of detection is considerably larger for OAs. Since they are still bright a few days after a GRB, in addition to a considerable probability of prepeak detection, they could also be observed with $\textit{Gaia}$'s second telescope and, if identified quickly, with ground-based telescopes.

\begin{figure}[!t]
\epsscale{1}
\plotone{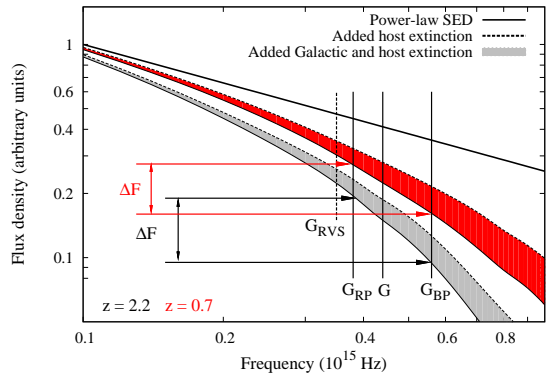}
\caption{Expected spectral energy distribution of detected GRB optical afterglows without extinction (solid line), with added host-galaxy extinction (dashed lines) and with additional Galactic extinction, where different values of extinction are considered (filled areas). Host-galaxy extinction was calculated at redshift $z = 0.7$ (red), corresponding to the average redshift of detected OAs in our simulation, and $z = 2.2$ (black), corresponding to the average observed redshift of GRBs, detected with the $\textit{Swift}$ satellite. Central frequencies of $\textit{Gaia}$'s photometric bands are  marked with vertical lines. See the text for more details.}
\label{fig10}
\end{figure}

Since the change in afterglow magnitude while being scanned by $\textit{Gaia}$ is not expected to be large enough to enable reliable way of identification of a transient source as a GRB optical afterglow, we consider other possibilities for GRB afterglow identification from one short detection. Since $\textit{Gaia}$ will have a photometric instrument onboard, there is the possibility to identify afterglow from its spectral energy distribution (SED). It has been observed that the SED of afterglows follows a power-law ($F_\nu  \propto \nu^{-\beta}$) modulated by extinction  \citep{schady, greiner}. We show the expected SED shape in UV-to-NIR frequency range in Figure \ref{fig10}, where an average spectral index value $\beta = 0.6$ has been assumed \citep{greiner}. As shown in many statistical analyses, in the majority of cases the extinction in GRB host galaxies is best described with a so-called SMC extinction profile \citep{pei}. Thus, we added SMC-type extinction to the SED (dashed lines in Figure \ref{fig10}), assuming $A_{\rm V} = 0.3$, which is the average determined extinction value in the $V$ band \citep{schady, greiner}. The extinction in the host-galaxy rest frame has been calculated for an average observed GRB redshift $z = 2.2$ \footnote{http://heasarc.gsfc.nasa.gov/docs/$\textit{Swift}$/archive/grb\_table/} and for redshift $z = 0.7$, which is the average redshift of detected OAs in our simulation. There is also a contribution to extinction by our Galaxy. We left the contribution of Galactic extinction to vary between $0 \leq A_{\rm V,GA} \leq 0.3$ and, using \citet{cardelli}, added Galactic contribution to the SED. In Figure \ref{fig10}, shaded areas mark the possible range of SED curves depending on the $A_{\rm V,GA}$ value. Vertical lines correspond to the central wavelengths of the $\textit{Gaia}$ photometric instrument's three wide passbands \citep{jordi}. Since $\textit{Gaia}$ photometry will be obtained by means of two low-resolution spectra (so-called red photometer (RP) and blue photometer (BP) spectra with central frequencies shown in Figure \ref{fig10} as $G_{\rm RP}$ and $G_{\rm BP}$ vertical lines), these spectra could provide us with enough information to determine the SED shape. In the case of bright sources, integrated flux from the radial velocity instrument ($G_{\rm RVS}$) can also be obtained. As shown in Figure \ref{fig10}, the change in flux density in the observational spectral range (marked as $\Delta F$) is expected to be large enough to be measurable. If the SED of a detected transient source resembles the one shown in Figure \ref{fig10}, the source could be considered to be a GRB afterglow. 

Another possibility to identify an on-axis afterglow is to cross-check the position and time of $\textit{Gaia}$ detection with GRB triggers by gamma-ray satellites (e.g., $\textit{Swift}$, Fermi, Integral). Ground-based follow-up observations would also be helpful in afterglow recognition, but their usefulness will critically depend on the delay between the $\textit{Gaia}$ detection and issuing of the alert to the community (which is expected to be about 24 hr) and the limiting magnitude of the follow-up campaign.

\section{Conclusions}
\label{s5}
We have shown that the number of expected GRB optical afterglow detections with $\textit{Gaia}$ satellite during its five years of expected operation is of the order of a few tens. Most of the expected detections are of the orphan afterglows, while the possibility of an on-axis afterglow detection is small (less than 10). The results are dependent on a set of various parameters, especially in the case of OAs. Since we made some assumptions and simplifications in the course of our simulation, we would like to point out that these results have to be interpreted as an order-of-magnitude estimations. In addition to the small number of expected detections, poor time sampling of the afterglow light curve and problem of afterglow identification make the study of GRB optical afterglows with $\textit{Gaia}$ difficult. On the other hand, since OAs have not been conclusively detected yet, a possible detection would provide valuable new data. 

\acknowledgments

AG acknowledges funding from the Slovenian Research Agency and from the Centre of Excellence for Space Science and Technologies SPACE-SI, an operation partly financed by the European Union, European Regional Development Fund, and Republic of Slovenia, Ministry of Higher Education, Science and Technology.

\appendix

\section{$\textit{Gaia}$'s scanning law}
\label{a1}
To describe $\textit{Gaia}$'s scanning law we first set aside its rotation around the Sun. Angles $\gamma$, $\Omega t$ and $\omega t$ (at an arbitrary time $t$) represent the well-known Euler angles \citep{hand}. Coordinates of the first and second telescope axes positions in the fixed coordinate system ($S$) with its origin at the satellite are described with $\textbf{r}_{1,2}$, while coordinates in the system of the scanning satellite ($S'$) are described with $\textbf{r'}_{1,2}$. Transformation between those two systems is given by matrix $U$ 
\begin{equation}
\label{eq12}
\textbf{r}_{1,2} = U\textbf{r'}_{1,2}, 
\end{equation}
which is obtained by multiplying three rotational matrices, each one describing a rotation with one of the Euler angles:
\begin{flushleft}
\footnotesize
\[
U =
\left( {\begin{array}{ccc}
\cos(\omega t)\cos (\Omega t) - \cos (\gamma) \sin (\Omega t) \sin (\omega t) & -\sin (\omega t) \cos (\Omega t) - \cos (\gamma) \sin (\Omega t) \cos (\omega t) & \sin (\gamma) \sin (\Omega t)\\
\cos(\omega t)\sin (\Omega t) + \cos (\gamma) \cos (\Omega t) \sin (\omega t) & -\sin (\omega t) \sin (\Omega t) + \cos (\gamma) \cos (\Omega t) \cos (\omega t) & -\sin (\gamma) \cos (\Omega t)\\
\sin (\gamma) \sin (\omega t) & \sin (\gamma) \cos (\omega t) & \cos (\gamma)
\end{array}} \right).
\]
\end{flushleft}
The first telescope axis is placed to be the $x'$ axis, i.e., $\textbf{r'}_{1} = (1, 0, 0)$. Considering the axes' separation angle $\beta$, the direction of the second telescope axis is described by $\textbf{r'}_{2} = (\cos \beta, -\sin \beta, 0)$. Now we introduce another (final) fixed coordinate system $\tilde{S}$, initially coinciding with $S$. The rotation around the Sun (for an angle depending on the time of the year) is obtained by rotating $S$ around the $x$ axis (Figure \ref{fig1}).

\end{document}